# Scaling Distributed All-Pairs Algorithms:
## Manage Computation and Limit Data Replication with Quorums

Cory J. Kleinheksel and Arun K. Somani

Department of Electrical and Computer Engineering
Iowa State University, Ames, Iowa 50011
cklein@iastate.edu; arun@iastate.edu

**Abstract.** In this paper we propose and prove that cyclic quorum sets can efficiently manage all-pairs computations and data replication. The quorums are O(N/√P) in size, up to 50% smaller than the dual N/√P array implementations, and significantly smaller than solutions requiring all data. Implementation evaluation demonstrated scalability on real datasets with a 7x speed up on 8 nodes with 1/3$^{rd}$ the memory usage per process.

The all-pairs problem requires all data elements to be paired with all other data elements. These all-pair problems occur in many science fields, which has led to their continued interest. Additionally, as datasets grow in size, new methods like these that can reduce memory footprints and distribute work equally across compute nodes will be demanded.

## 1   Introduction

In elementary schools and introductory computer science courses a popular "handshake" problem [1] is often taught and it goes something like this: $P$ people attend a party and a popular greeting is to shake hands, how many handshakes take place? After discussion and manipulation the answer of $\binom{P}{2} = \frac{P(P-1)}{2}$ is derived.

This "handshake" problem is not reserved for the teaching introductory topics. In databases this manifests as a self-join without a join condition, forcing all tuples to interact with all other tuples. In physics, the n-body problem predicts the position and motion of $n$ bodies by calculating the total forces every body has on every other body. In biometrics applications, a similarity matrix can be formed using a set of images compared with itself using facial recognition [2]. In metagenomics, finding a protein's likeness to every other protein is a crucial part of forming the complex graphs used in protein clustering, which has led to new discoveries of protein functions [3].

### 1.1   Acceleration of Applications

Accelerating the execution of many of these important applications has been done using multicore CPUs, FPGAs, GPUs, Intel's many-core MIC, and distributed clusters. In [4] the authors provide a generalized framework to solve these all-pair classification of algorithms and show performance improvements for biometrics and data mining applications in a distributed system, e.g., cloud. A different approach was taken for a bioinformatics application seeking to reconstruct gene co-expression networks. The PCIT algorithm [5] was chosen to identify significant gene correlations. This method was optimized for Intel's multicore Xeon and many-core MIC [6].

Every element interacting with every other element leads to a natural result of having all elements present in memory. The generalized framework [4] showed that efficiently distributing all of the input data to all of the nodes prior to beginning execution resulted in faster turnaround times than reading from the disk on demand. The optimization of the PCIT algorithm [6] experienced needing all of the data in memory and created a second optimization strategy with longer runtimes, but had a minimal memory usage footprint.

### 1.2  Relaxing the All Elements Present Requirement

N-body problems have a natural all-pairs decomposition called atom-decomposition [7] that is based on equal distribution of $N$ element responsibilities to $P$ parallel processes. To address load imbalances and the need to communicate all data to all processes, the authors proposed a method to perform force-decomposition which still requires input data replication, but reduced it to 2 arrays of size $\frac{N}{\sqrt{P}}$ elements per process. The authors in [8] showed that data replication in the system can be variable ($c$); and when $c = \sqrt{P}$, a lower bound on communication is achieved. When $c = 1$, their solution behaved similar to atom-decomposition, although requiring only 2 arrays of N/P elements per process. When $c = \sqrt{P}$, their solution behaved similar to force-decomposition and still required 2 arrays of size $\frac{N}{\sqrt{P}}$ elements per process.

Minimizing the amount of data replication in a distributed system, while maintaining efficient all-pairs algorithm operation, is a recurring theme in this classification of algorithms. Quorum systems are commonly used for coordination and mutual exclusion in distributed systems [9], [10]. Their decentralized approach and slow quorum growth rate compared to the system size are two of the reasons that make them a good tool in managing replicated data [11]. In 1985, quorums of size $O(\sqrt{P})$ were proven using finite projective planes [12]. Relaxed difference sets later were used to create size $O(\sqrt{P})$ cyclic quorum sets [10].

### 1.3  Scaling All-Pairs Algorithms

In this paper, we utilize the slow quorum growth rate compared to number of processes to scale the all-pairs classification of algorithms. We provide a proof that cyclic quorum sets have an all-pairs property that allows our solution to use a single array of size $O\left(\frac{N}{\sqrt{P}}\right)$ elements per process, significantly less than current solutions that have $N$ elements per process and up to 50% improvement over those that would require two arrays of size $\frac{N}{\sqrt{P}}$ elements per process. These cyclic quorums are optimal for all Singer difference sets [13] and near-optimal for all others. For processes $P = 4, \ldots, 111$, our work uses the optimal cyclic quorums from [10].

The rest of the paper is organized as follows: Section 2 formalizes the all-pairs problem. Section 3 quorum sets, and more specifically cyclical quorum sets, are defined. Section 4 provides a definition of the all-pairs property and that cyclic quorums satisfy the property. Lastly, in Section 5 we experiment with a bioinformatics all-pairs application to show the scalability and memory efficiency of our all-pairs quorum methods.

## 2 All-Pairs Problem

The all-pairs problem (or "handshake" problem) occurs in many different fields and occurs in a broad classification of algorithms. On the surface the problem is very straight forward as shown in Figure 1. Given a set of data elements (seven in our example), an algorithm pairs all elements with all other elements. Notice that it is not necessary to explicitly form a $(D_2, D_1)$ pair because the pair can be formed by the $(D_1, D_2)$ pair already present.

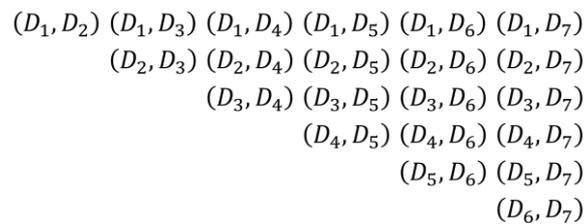

Fig. 1. All-pairs of seven data elements.

### 2.1 General All-Pairs Problem Definition

The pseudocode for a general all-pairs algorithm would look like the following:

```
Given: Array D
for i ← 1, length(D)-1 do
  for j ← i+1, length(D) do
    perform work on pair (i,j)
```

Stated more formally:

$$\text{Set of } N \text{ elements } D = \{d_1, d_2, \ldots, d_N\} \tag{1}$$

$$Pair(d_i, d_j), \text{ where } 1 \leq i < N \text{ and } i < j \leq N \tag{2}$$

Equation 1 enumerates the data elements being paired, while Equation 2 performs all data pairings resulting in $\binom{N}{2} = \frac{N(N-1)}{2}$ element pairings.

### 2.2 Distributed All-Pairs Problem Definition

The distributed all-pairs problem distributes the $\binom{N}{2}$ element pairings across $P$ processes. The $N$ data elements are grouped into $P$ datasets:

$$\text{Set of } P \text{ datasets } \widehat{D} = \{D_1, D_2, \ldots, D_P\} \tag{3}$$

$$D_i \subseteq D, \ i \in 1, 2, \ldots, P \tag{4}$$

$$D = \bigcup_{i=1}^{P} D_i \tag{5}$$

Equation 3 enumerates a set of datasets, one dataset per process. Equation 4 states that each dataset is equal to or a subset of the complete dataset $D$ and the union of those subsets equals the global dataset $D$ (Eq. 5). Methods of distributing the work and data vary, e.g., [4], [7], [8]. Some implementations give all of $D$ to all processes and each is responsible for a different portion of the $\binom{N}{2}$ element pairings. Other implementations have mechanisms to generate different data subsets and then compute the global pairing of all of $D$ by pairing individual subsets.

$$Pair(D_i, D_j), \text{ where } 1 \leq i < P \text{ and } i \leq j \leq P \tag{6}$$

Equation 6 performs the pairing of datasets $D_i$ and $D_j$, whereas Equation 2 is preforming pairing of particular data elements $d_i$ and $d_j$. Additionally, in Equation 6 the range of index $j$ is altered to allow for the pairing of $D_i$ with itself. This is unnecessary in Equation 2 because elements in general would not need to be paired with themselves; however once placed in a subset, it is still necessary that elements within a subset be paired with others within the same subset and not simply with only other subsets.

In the next section, we introduce the quorum sets that we use in our work to equally distribute work and data for the distributed all-pairs problem.

## 3  Quorum Sets

In distributed communication and algorithms, coordination, mutual exclusion, data replication and consensus implementations have grouped $P$ processes or nodes into sets called quorums [11]. This organization can minimize communications in operations like negotiating access to a global resource or reaching a joint, distributed decision.

A quorum set minimally has the property that all quorums must intersect. Specifically for distributed implementations, it is also desirable that each quorum have equal work and equal responsibility within the quorum set. Not every grouping of nodes into sets (quorums) will result in having these three properties, nor will the quorum sizes be minimal. [12] proved the lower bound on the size of a quorum set with these three properties. Cyclic quorum sets have these properties and are proposed in this paper for efficient all-pairs problem computation and data replication management.

### 3.1 Defining Quorum Sets

From Equation 3, $\widehat{D}$ is set of datasets, one for each of the $P$ processes. Sets $S_i$ are subsets of $\widehat{D}$ (Eq. 7). When set $Q$ of subsets (Eq. 8) covers all of $\widehat{D}$ (Eq. 9) and all subsets also have non-empty intersections (Eq. 10), then set $Q$ is called a quorum set.

$$S_i \subseteq \widehat{D},\ i \in 1, 2, \ldots, P \tag{7}$$

$$Q = \{S_1, \ldots, S_P\} \tag{8}$$

$$\cup_{i=1}^{P} S_i = \{D_1, \ldots, D_P\} = \widehat{D} \tag{9}$$

$$S_i \cap S_j \neq \emptyset,\ \forall\, i, j \in 1, 2, \ldots, P \tag{10}$$

The lower bounds for the maximum individual quorum size (i.e., $|S_i|$) in a minimum set is $k$, where Equation 11 holds and $(k-1)$ is a power of a prime, proved through equivalence to finding a finite projective plane [12]. Additionally, it is desirable that each quorum $S_i$ in the quorum set be of equal size (Eq. 12), such that there is equal work and it is desirable that each dataset $D_i$ be contained in the same number of quorums (Eq. 13), such that there is equal responsibility.

$$P \leq k(k-1) + 1 \tag{11}$$

$$|S_i| = k,\ \forall i \in 1, 2, \ldots, P \tag{12}$$

$$D_i \text{ is contained in } k\ S_j\text{'s},\ \forall i \in 1, 2, \ldots, P \tag{13}$$

### 3.2 Defining Cyclic Quorum Sets

Cyclic quorum sets are based on cyclic block design and cyclic difference sets, however searching for optimal sets requires an exhaustive search [10]. Cyclic quorum sets are unique in that once the first quorum (Eq. 14) is defined the remaining quorums in the set can be generated via incrementing the dataset indices (modulus to keep dataset indices within bounds is not shown in Equation 15 for conciseness). For simplicity, assume $D_1 \in S_1$ without loss of generality (any one-to-one re-mapping of dataset indices can result in this assumption).

$$S_1 = \{D_1, \ldots, D_j\} \tag{14}$$

$$S_i = \{D_{1+(i-1)}, \ldots, D_{j+(i-1)}\} \tag{15}$$

For our work, we used the $P = 4, \ldots, 111$ optimal cyclic quorums from [10]. In the next section, we define and prove that cyclic quorum sets have an all-pairs property that makes them ideal for managing the distributed all-pairs problem.

# 4 All-Pairs Property for Quorum Sets

Cyclical quorums were introduced in the previous section as having a small size and equal work/responsibility properties. However, on the surface it is not apparent how these small/equable cyclic quorum sets can support the pairing of all $\{D_1, \ldots, D_P\}$ distributed datasets to solve the general all-pairs problem (i.e., Section 2, Equation 6). In this section we define the all-pairs property for quorum sets and provide a proof that cyclical quorum sets satisfy this property.

## 4.1 All-Pairs Property

As all-pairs algorithms scale using multiple processes and distribute the $\binom{N}{2}$ work, it remains necessary that all pairs of elements are present in at least one process's memory. When using quorums in the distributed system, we assign each process $i$ a quorum $S_i$ of datasets. It is from there that we define the all-pairs property for quorums:

$$\exists S_i \ni (D_j, D_k) \quad \forall j, k \in 1, 2, \ldots, P, \quad where\ S_i \in Q \tag{16}$$

Equation 16 states that for every pairing of datasets in $\widehat{D}$ there exists at least one quorum in the quorum set $Q$ that contains the pair. Quorum sets with this all-pairs property can be used to satisfy Equation 6 from Section 2, that defined all of the work that needed to be performed in the distributed all-pairs problem.

## 4.2 Cyclic Quorums have the All-Pairs Property

We use the relationship between cyclic quorum sets and difference sets [10] as part of a proof that the all-pairs property is satisfied.

*Definition 1.* Set $A = \{a_1, \ldots, a_k\}$ $modulus\ P$, $a_i \in 0, \ldots, P-1$ is a relaxed $(P, k) - difference\ set$ if for every $d \neq 0\ modulus\ P$, $\exists (a_i, a_j)$, $a_i, a_j \in A$ such that $a_i - a_j = d\ modulus\ P$.

Definition 1 defines a relaxed difference set as a set of integers whose values are greater than or equal 0 and less than $P$. It has a restriction that every integer from 0 to $(P - 1)$ must also be able to be formed from the difference of some pair of integers in the set (using modulus when necessary.)

*Definition 2.* The cyclic quorum set $Q$ defined by set $S_i = \{a_1 + (i - 1), \ldots, a_k + (i - 1)\}\ modulus\ P$, $i \in 0, \ldots, P - 1$ is a relaxed $(P, k) - difference\ set$ $A = \{a_1, \ldots, a_k\}\ modulus\ P$, $a_i \in 0, \ldots, P - 1$.

The intuition for Definition 2 relies on the quorum set's intersection property, $S_i \cap S_j \neq \emptyset$, $\forall\ i, j$ (Eq. 10). By contradiction, assume that set $A$ was not a relaxed difference set, then there would be value $v \neq 0\ modulus\ P$ that no difference $(a_i - a_j)$, $a_i, a_j \in A$ equaled. Given that every quorum intersects in the set $Q$, there must be a shared item in $S_0$ and $S_v$. Equation 17 assumes the shared item is at indices $l$ and $m$ respectively, hence the shared item $s_{0,l}$ and $s_{v,m}$ are differenced on the left-hand side. Using the quorum set definition, the values for the items are substituted on the right side. Equation 18 uses the quorum intersection to simplify the left side to 0 before rebalancing to show that the assumption is false, hence set $A$ is a relaxed difference set.

$$s_{0,l} - s_{v,m} = \bigl(a_l + (0-1)\bigr) - \bigl(a_m + (v-1)\bigr) \; modulus \; P \tag{17}$$

$$a_l - a_m = v \; modulus \; P \tag{18}$$

*Theorem 1.* The cyclic quorum set $Q$ defined by set $S_i = \{a_1 + (i-1), \ldots, a_k + (i-1)\} \; modulus \; P, \; i \in 0, \ldots, P-1$ satisfies the all-pairs property (Section 4.1)

Proof by contradiction: Assume that the all-pairs property is not satisfied. Then there must a pair of integers $(a_x, a_y)$, $a_x, a_y \in 0, \ldots, P-1$ that are not present together in any quorum $S_i \in Q$. Integers $(a_x, a_y)$ have the following differences:

$$(a_x - a_y) \; modulus \; P \; \text{and} \; (a_y - a_x) \; modulus \; P. \tag{19}$$

Definition 2 states both differences in Equation 19 are formed at least once from the difference set $A = \{a_1, \ldots, a_k\}$. Assume that integers $(a_l, a_m)$, $a_l, a_m \in A$ form those specific differences:

$$(a_l - a_m) \; modulus \; P \; \text{and} \; (a_m - a_l) \; modulus \; P. \tag{20}$$

$$(a_l - a_m) \; modulus \; P = \bigl(a_l + (i-1)\bigr) - \bigl(a_m + (i-1)\bigr) \; modulus \; P \tag{21}$$

Using the cyclic quorum set definition and distributive property of modular arithmetic, Equation 21 shows all $S_i$ cyclic quorums can form the same differences. Using this result and that Equation 20 must produce the differences from Equation 19 otherwise Definition 2 would be false, we can now combine Equations 19-21.

$$(a_x - a_y) \; modulus \; P = (a_l - a_m) \; modulus \; P, \; \text{or} \tag{22}$$

$$(a_x - a_y) \; modulus \; P = (a_m - a_l) \; modulus \; P \tag{23}$$

$$(a_x - a_y) \; modulus \; P = \bigl(a_l + (i-1)\bigr) - \bigl(a_m + (i-1)\bigr) \; modulus \; P \tag{24}$$

Equations 22 and 23 are not necessarily both true at the same time, rather they enumerate the two values that the differences can take. $(a_y - a_x)$ is not shown as it is just the inverse of the two equations shown. Equation 24 can be seen as combining all four into a single statement. All combinations of $(a_x - a_y)$ differences can be expressed in terms of difference set $A$ and a corresponding quorum integer $i$. This leads to our final result by distributing the modulus and separating the terms.

$$a_x \bmod P = \big(a_l + (i-1)\big) \bmod P \tag{25}$$

$$a_y \bmod P = \big(a_m + (i-1)\big) \bmod P \tag{26}$$

Equations 25 and 26 show that integers $(a_x, a_y)$, $a_x, a_y \in {0, ..., P-1}$ are present together in a quorum defined by difference set $A$ and integer $i \bmod P$. This contradicts the assumption, hence cyclic quorum sets do satisfy the all-pair property.

## 5   Experimental Application Results

To evaluate the performance of our cyclic quorum set method, we modified an existing all-pairs application [6] to scale to larger datasets and at the same time be able to utilize more resources. The algorithm implemented the distributed all-pairs problem defined in Equation 6 using the cyclical quorum sets defined in Section 3.2.

### 5.1   PCIT Algorithm

The partial correlation coefficients combined with an information theory approach (PCIT) algorithm was introduced by [5]. The algorithm can be used for gene co-expression network reconstruction and help to identify novel biological regulators. The technique processes $N$ genes by building an $O(N^2)$ matrix and using a guilt-by-association heuristic to analyze node pair partial correlations identifying whether a gene expression correlation is or is not meaningful.

### 5.2   Results

Testing was conducted using Cyence, an HPC machine at Iowa State University. Every node has dual 8-core processors and 128GB of memory. Two real and one synthetic input dataset were utilized with up to 20 execution runs per test. The single node PCIT algorithm from [6] was run with 16 OpenMP threads on a node by itself. The quorum implementations ran with one MPI process per processor (two per node) and 8 OpenMP threads per process.

Figure 2 shows a 7x speed up in computation performance using 8 nodes as well as over 2/3$^{rd}$ reduction of memory per process due to our cyclic quorum method for the all-pairs problem. On the left side of each graph is the single node optimized PCIT algorithm performance. For comparison, this single node performance is then divided by the number of additional nodes utilized by the quorum implementation and displayed in a decreasing curved line respective to each input file. As the number of nodes increase and the memory per process continues to decrease, the performance meets or exceeds the corresponding ideal lines. However, scaling only to two nodes encountered suboptimal and inconsistent performance as can be seen in the vertical bar extending above the curved ideal scaling line and the larger 95% confidence error bar that can be seen.

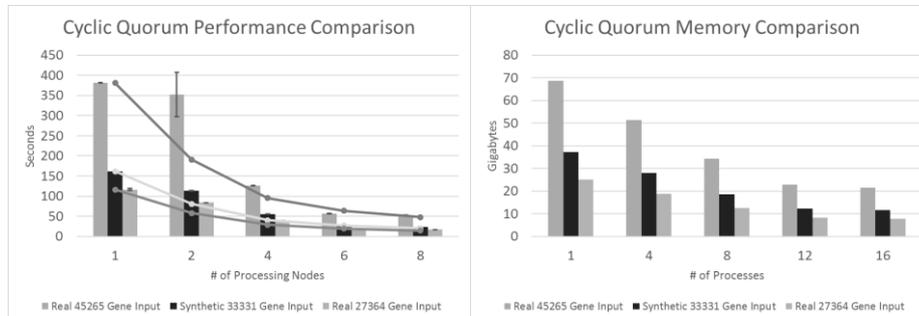

**Fig. 2.** Performance and memory comparison between original single node algorithm and our cyclic quorum set implementation. Three inputs of different sizes were used. The figure on the left has the algorithm performance on the vertical bars and ideal scaling from one node to many illustrated with the curved horizontal lines. The figure on the right shows the memory requirement reduction per process as the application is scaled across more resources.

## 6   Conclusions

In this paper we proposed and proved that cyclic quorum sets are suitable for the all-pairs problem, sometimes known as "handshake" problems. The dataset quorums are $O\left(\frac{N}{\sqrt{P}}\right)$ in size, up to 50% smaller than the dual $\frac{N}{\sqrt{P}}$ array implementations, and significantly smaller than solutions requiring all data. Algorithm design can be simplified as all of the data needed for pairing exists in a process's dataset quorum.

Implementation evaluation took a single node bioinformatics all-pairs implementation and demonstrated scalability on real datasets. Computation had a 7x speed up and memory usage per process was cut by 2/3$^{rd}$ when using eight nodes.

Future work includes investigating optimization opportunities, particularly demonstrating the efficiency and power of utilizing Singer difference sets and achieving efficient performance even for non-Singer difference sets. The need for this work is motivated by the four process test that performed suboptimal in Figure 2. Lastly, for applications where redundancy is important, we are investigating using quorum redundancy to deliver memory and computationally efficient solutions.

**Acknowledgements.** Research funded in part by NSF Graduate Research Fellowship Program, IBM Ph.D. Fellowship Program, Symbi GK-12 and Trinect Fellowships at Iowa State University, and the Jerry R. Junkins Endowment at Iowa State University. The research reported in this paper is partially supported by the HPC@ISU equipment at Iowa State University, some of which has been purchased through funding provided by NSF under MRI grant number CNS 1229081 and CRI grant number 1205413. Any opinions, findings, and conclusions or recommendations expressed in this material are those of the author(s) and do not necessarily reflect the views of the funding agencies.